\documentclass[a4paper,aps,prd,10pt,preprintnumbers,showpacs,twocolumn,superscriptaddress,nofootinbib,amsmath,amssymb]{revtex4-1}
\usepackage[dvips]{graphics}

\def\im#1{Im(#1)}

\newcommand{\ie}{{i.e.,}~}

\newcommand{\rem}[1]{}

\begin{document}
\title{Quasinormal modes and a new instability of Einstein-Gauss-Bonnet black holes in the de Sitter world}

\author{M. A. Cuyubamba}
\email{marco.espinoza@ufabc.edu.br}
\affiliation{Centro de Matem\'atica, Computa\c{c}\~ao e Cogni\c{c}\~ao,
  Universidade Federal do ABC (UFABC), Rua Aboli\c{c}\~ao, CEP:
  09210-180, Santo Andr\'e, SP, Brazil}

\author{R. A. Konoplya}
\email{konoplya\_roma@yahoo.com}
\affiliation{Theoretical Astrophysics, Eberhard-Karls University of T\"ubingen, T\"ubingen 72076, Germany}

\author{A. Zhidenko}
\email{olexandr.zhydenko@ufabc.edu.br}
\affiliation{Centro de Matem\'atica, Computa\c{c}\~ao e Cogni\c{c}\~ao,
  Universidade Federal do ABC (UFABC), Rua Aboli\c{c}\~ao, CEP:
  09210-180, Santo Andr\'e, SP, Brazil}

\begin{abstract}
  Analysis of time-domain profiles for gravitational perturbations shows that Gauss-Bonnet black holes in a de Sitter world possess a new kind of dynamical instability which does not take place for asymptotically flat Einstein-Gauss-Bonnet black holes. The new instability is in the gravitational perturbations of the scalar type and is due to the nonvanishing cosmological constant. Analysis of the quasinormal spectrum in the stability sector shows that although the scalar type of gravitational perturbations alone does not obey Hod's conjectural bound, connecting the damping rate and the Hawking temperature, the vector and tensor types (and thereby the gravitational spectrum as a whole) do obey it.
\end{abstract}
\maketitle

\section{Introduction}

Higher-curvature corrected theories of gravity are an interesting alternative to Einstein gravity. In higher than four-dimensional spacetimes, the low-energy limit of the heterotic string theory predicts the Gauss-Bonnet-type (second order in curvature) correction to the Einstein action. The Einstein-Gauss-Bonnet theory shows a number of qualitatively new interesting
features. First, unlike Schwarzschild or Kerr black holes, small black holes in the Einstein-Gauss-Bonnet theory are unstable against small gravitational perturbations \cite{GleiserDotti}. The graviton can experience time advance when Gauss-Bonnet corrections are taken into consideration \cite{Camanho}. The anti-de Sitter spacetime, which is unstable against nonlinear perturbations in Einstein gravity \cite{Bizon}, restores stability in the Gauss-Bonnet theory \cite{Deppe}. Finally, Hawking radiation of even softly Gauss-Bonnet-corrected black holes occurs at the evaporation rate which is many orders slower than that of the black hole in the higher-dimensional Einstein theory \cite{Konoplya:2010vz}.

Special attention  in the literature is devoted to the issue of gravitational stability of black holes (see reviews in \cite{Konoplya:2011qq} and references therein), because stability is a necessary criterium of viability of a black hole's model. Analytical analysis of black holes' (in)stability is not an easy task even for relatively simple black-hole solutions, so that for the most of cases numerical treatment through consideration of the black hole oscillation (\emph{quasinormal}) spectrum \cite{Kokkotas:1999bd} comes into play. The quasinormal modes of black holes were extensively studied by theorists during the past years. Now there is a strong indication that quasinormal modes have been recently experimentally observed in the gravitational-wave signal from an event that might be a merger of binary black-hole system \cite{Abbott:2016blz,TheLIGOScientific:2016src}. In \cite{Abbott:2016blz,TheLIGOScientific:2016src}, it was shown that the gravitational-wave signal is consistent with the Einstein theory of gravity. At the same time, the binary black hole evolution was simulated by LIGO and VIRGO collaborations only within the Einstein theory, leaving the possible consistency of alternative  theories to further investigation \cite{TheLIGOScientific:2016src}. Simple  estimations of orders show that rather large indeterminacy of the black hole's parameters should leave the window for alternative theories of gravity open \cite{Konoplya:2016pmh}.

Being an essential criterium for four-dimensional asymptotically flat or de Sitter spacetimes, the (in)stability of black holes in $(D>4)$-dimensional asymptotically de Sitter and anti-de Sitter spacetimes gains additional interest due to the renown gauge-gravity duality, which can be formulated either in terms of AdS-CFT \cite{Maldacena:1997re} or dS-CFT \cite{Strominger:2001pn} correspondences.

The (in)stability of asymptotically flat Einstein-Gauss-Bonnet black holes was studied in \cite{GleiserDotti} where it was shown that the small black holes are unstable against  gravitational perturbations for $D=5$ and $D=6$. The time-domain picture of evolving of the instability showed that, unlike other black holes' instabilities, it develops at relatively late time after a long period of damped quasinormal oscillations \cite{Konoplya:2010vz}.  Further study of the black holes' instability for the Lovelock generalization \cite{Lovelock:1971} of Gauss-Bonnet gravity was performed in \cite{Takahashi:2010ye}, while quasinormal modes were studied in \cite{Abdalla:2005hu,Yoshida:2015vua,Moura:2006pz}.

Here we shall make the next step and analyze (in)stability of black holes in Gauss-Bonnet gravity, allowing for the positive, nonzero cosmological constant and corresponding to the de Sitter asymptotic\footnote{Notice, that the Gauss-Bonnet equations with a positive $\Lambda$-term have also nonasymptotically de Sitter solutions, which are not considered here.}. Higher-dimensional analogue of the Schwarzschild-de Sitter black hole (given by the Tangherlini-de Sitter metric) is stable in Einstein gravity \cite{Konoplya:2007jv}. Here we shall show that this is not so for the Gauss-Bonnet theory. Looking at the already known instability of asymptotically flat small Gauss-Bonnet black holes, instability of Gauss-Bonnet-de Sitter solution could also be expected at least in some region of parameters. Here we shall show that in addition to this expected instability, there is a new type of instability which occurs owing to the nonzero cosmological constant and does not take place for asymptotically flat higher-dimensional Gauss-Bonnet or Schwarzschild (Tangherlini) black holes. We suppose that this instability, triggered by the $\Lambda$-term in the Gauss-Bonnet theory, might have the same origin as the instability of higher-dimensional Reissner-Nordstr\"om-de Sitter black holes, which was found in \cite{Konoplya:2008au} and later studied for the regime of large \cite{Tanabe:2015isb} and arbitrary \cite{Konoplya:2013sba} numbers of spacetime dimensions $D$. Here, through the thorough study of the quasinormal spectrum of gravitational perturbations, we shall find the parametric regions of instabilities of the Einstein-Gauss-Bonnet-de Sitter black holes in various spacetime dimensions.

In addition to the stability study, there is another motivation to study quasinormal modes of Gauss-Bonnet-de Sitter black holes. In \cite{Hod:2006jw}, S. Hod made a proposal stating that  there exists a universal bound on the relaxation rate. According to Hod's proposal, in the quasinormal spectrum of any black hole there is always a mode whose damping rate (given by the absolute value of the imaginary part of the quasinormal frequency $|\im\omega|$) is not higher than $\pi T_{BH}$, where $T_{BH}$ is the Hawking temperature of the black hole.
Hod's inequality is satisfied for the asymptotically flat Gauss-Bonnet (GB) as well as for Schwarzschild-de Sitter black holes (though for the GB case this check seems never been published). Therefore, the natural question would be -- whether this inequality is satisfied also for the Gauss-Bonnet-de Sitter solution. Here we shall show that the scalar type of gravitational perturbations has the fundamental (lowest damping) mode which is above Hod's minimum. At the same time the other two (vector and tensor) types of perturbations satisfy the inequality. We shall argue that this rather signifies the nonviolation of Hod's proposal by the Gauss-Bonnet-de Sitter black holes.

The paper is organized as follows. In section~\ref{sec:ft}, the main information about the Einstein-Gauss-Bonnet-de Sitter background is given. Section~\ref{sec:grav} briefly reviews the perturbation equations and the main properties of the master wave equations. Section~\ref{sec:timeint} is devoted to description of the numerical time-domain integration method which we used for analysis of evolution of perturbations. Section~\ref{sec:res} relates the obtained results and the found regions of instability. Section~\ref{sec:Hod} discusses the obtained data for quasinormal modes and the validity of Hod's proposal. In the Conclusion, we sketch the obtained results and ongoing work in this direction.

\section{Higher-curvature corrected gravity and black hole solutions}\label{sec:ft}

The natural generalization of the Einstein theory of gravity to the $(D>4)$-dimensional spacetime was carried out by Lovelock \cite{Lovelock:1971}. The Lagrangian has the following form,

\begin{eqnarray}
  \mathcal{L} &=& \sum_{m=0}^kc_m\mathcal{L}_m,\label{gbg1} \\
  \mathcal{L}_m &=& \frac{1}{2^m}\delta^{\mu_1\nu_1 \ldots\mu_m\nu_m}_{\lambda_1\sigma_1\ldots\lambda_m\sigma_m}\,R_{\mu_1\nu_1}^{\phantom{\mu_1\nu_1}\lambda_1\sigma_1} \ldots R_{\mu_m\nu_m}^{\phantom{\mu_m\nu_m}\lambda_m\sigma_m},\label{gbg2}
\end{eqnarray}
where $\delta^{\mu_1\nu_1\ldots\mu_m\nu_m}_{\lambda_1\sigma_1\ldots\lambda_m\sigma_m}$ and $R_{\mu\nu}^{\phantom{{\mu\nu}}\lambda\sigma}$ are the $D$-dimensional Kronecker delta and Riemann tensors, $k=\left[{(D-1)/2}\right]$, and $c_m$ are arbitrary constants of the theory.

Here we shall be limited by the well-known Gauss-Bonnet theory, that is the Lovelock theory truncated at the second-order term in the curvature. We use geometrized units $16\pi G=1=\hbar$ and define $c_1=1$, $c_0=-2\Lambda$, where $\Lambda$ is the cosmological constant, and $c_2=\alpha/2$ \cite{Takahashi:2010ye}. This limit describes the low-energy regime of the heterotic string theory. The Einstein-Gauss-Bonnet Lagrangian takes the following form
\begin{equation}\label{gbg3}
  \mathcal{L}=-2\Lambda+R+\alpha(R_{\mu\nu\lambda\sigma}R^{\mu\nu\lambda\sigma}-4\,R_{\mu\nu}R^{\mu\nu}+R^2),
\end{equation}
where $\alpha=1/2\pi\ell_s^2$ is a positive coupling constant.

As part of the Lovelock theory, Gauss-Bonnet gravity is self-consistent in five and six dimensions. In a more general context, which includes string theory and gauge/gravity duality, we do not need to be always limited by the $D=5, 6$ cases only. Thus, although we shall concentrate on five- and six-dimensional cases, part of our computations will include higher than six-dimensional black holes.

An exact static vacuum solution of the Einstein-Gauss-Bonnet equations can be written in the form
\begin{equation}\label{gbg4}
  ds^2=-f(r)dt^2+\frac{1}{f(r)}dr^2 + r^2\,d\Omega_n^2
\end{equation}
where $d\Omega_n^2$ is the $(n=D-2)$-dimensional line element of $S^n_\kappa$ manifold of constant curvature $\kappa=\pm 1,0$ and
\begin{equation}
f(r)=\kappa-r^2\,\psi(r),
\end{equation}
where $\psi(r)$ satisfies
\begin{equation}
W[\psi]\equiv\frac{\alpha n (n - 1) (n - 2)}{4} \psi^2 + \frac{n}{2}\psi - \frac{\Lambda}{n + 1} = \frac{\mu}{r^{n + 1}}\,,
\end{equation}
where $\mu$ is a positive constant proportional to the black-hole mass.

This quadratic equation has two solutions, given by
\begin{eqnarray}\label{psi}
&&\psi(r)=\frac{1}{\alpha(n-1)(n-2)}\times\\\nonumber&&\times
\left(\epsilon\sqrt{1+\frac{4\alpha (n-1)(n-2)}{n}\left(\frac{\mu}{r^{n+1}}+\frac{\Lambda}{n+1}\right)}-1\right)\,,
\end{eqnarray}
where $\epsilon=\pm1$. In this paper we shall study only the asymptotically de Sitter solutions, \ie when $\epsilon=1$ and $\Lambda>0$. Various properties of this metric were, in particular, studied in \cite{Cai:2001dz}.

It was observed in \cite{Konoplya:2008ix} that the accurate computation of $\psi(r)$ requires higher precision for the arithmetic operations. In order to decrease the relative error in (\ref{psi}) we use its alternative equivalent form (for $\epsilon=1$)
\begin{equation}\label{gbg6}
  \psi(r)=\frac{4\left(\frac{\mu}{r^{n+1}}+\frac{\Lambda}{n+1}\right)}{n+\sqrt{n^2+4\alpha n(n-1)(n-2)\left(\frac{\mu}{r^{n+1}}+\frac{\Lambda}{n+1}\right)}},
\end{equation}
which apparently allows us to perform all the computations using standard 32-bits floating-point arithmetics.

Furthermore, the $\alpha\rightarrow 0$ limit of Gauss-Bonnet gravity leads to the higher-dimensional solution of the Einstein theory -- Schwarzschild-Tangherlini spacetime \cite{Tangherlini:1963bw}
\begin{equation}\label{gbg7}
  f(r)=\kappa-\frac{2r^2}{n}\left(\frac{\mu}{r^{n+1}}+\frac{\Lambda}{n+1}\right).
\end{equation}

For $\kappa=1$ we have a compact (spherical) black hole with the event horizon radius $r_H$, which corresponds to the smallest positive root of the equation $f(r)=0$. Measured in units of length for any value of $D$, this quantity is convenient for parametrization of the black hole mass, which can be expressed as
\begin{equation}\label{nr1}
  \mu=\frac{n\,r_H^{n-1}}{4}\left(2+\frac{\alpha(n-2)(n-1)}{r_H^2}-\frac{4\Lambda  r_H^2}{n(n+1)}\right),
\end{equation}
In the de Sitter spacetimes the span of the spatial coordinate is limited by the cosmological horizon $r_C>r_H$, which we use in order to parametrize the cosmological constant as
\begin{eqnarray}\label{cosmological}
  \Lambda&=&\frac{n(n+1)}{2}\Biggr(\frac{r_C^{n-1}-r_H^{n-1}}{r_C^{n+1}-r_H^{n+1}}\\\nonumber&&\qquad+\frac{\alpha(n-1)(n-2)}{2}\frac{r_C^{n-3}-r_H^{n-3}}{r_C^{n+1}-r_H^{n+1}}\Biggr),
\end{eqnarray}

In the limit $r_C\rightarrow r_H$ we obtain the extremal value of the cosmological constant, which is given as follows
\begin{equation}\label{nr2}
  \Lambda_{extr}=\frac{n(n-1)}{2\,r_H^4}\left(r_H^2+\frac{(n-2)(n-3)}{2}\alpha\right).
\end{equation}
Limit $r_C\rightarrow\infty$ corresponds to the asymptotically flat spacetime ($\Lambda=0$).

Hereafter, we measure all the quantities in units of the event horizon, \ie we introduce dimensionless parameters, $0\leq\frac{r_H}{r_C}<1$ and $\frac{\alpha}{r_H^2}\geq0$, while frequencies are measured in the units of inverse horizon radius $r_H^{-1}$.

\section{Gravitational perturbations and the effective potentials}\label{sec:grav}

The linear perturbations $h_{\mu\nu}$ around the background $g_{\mu\nu}$ (\ref{gbg4}) can be written as
\begin{equation}\label{gp1}
  g_{\mu\nu}\rightarrow g_{\mu\nu}+h_{\mu\nu},\qquad\quad |h_{\mu\nu}|\ll|g_{\mu\nu}|.
\end{equation}

Taking the variation of the Einstein-Gauss-Bonnet equations for vacuum solutions,
\begin{equation}\label{gp2}
  \delta G_{\mu}^{\phantom{\mu}\nu}=\Lambda\,\delta G_{(0)\mu}^{\phantom{(0)\mu}\nu}+\delta G_{(1)\mu}^{\phantom{(1)\mu}\nu}+\alpha\, \delta G_{(2)\mu}^{\phantom{(2)\mu}\nu}=0,
\end{equation}
where
\begin{eqnarray}
  G_{(0)\mu}^{\phantom{(0)\mu}\nu} &=& \delta_\mu^\nu,\label{gp3} \\
  G_{(1)\mu}^{\phantom{(1)\mu}\nu} &=& R_\mu^{\phantom{\mu}\nu}-\frac{1}{2}\delta_\mu^\nu\,R\label{gp4}
\end{eqnarray}
and
\begin{eqnarray}
  G_{(2)\mu}^{\phantom{(2)\mu}\nu} &=& R_{\lambda\mu}^{\phantom{\lambda\mu}\delta\sigma}R_{\delta\sigma}^{\phantom{\delta e}\lambda\nu}-2R_\delta^{\phantom{\delta}\lambda}R_{\lambda\mu}^{\phantom{\lambda\mu}\delta\nu} -2R_\mu^{\phantom{\mu}\lambda}R_\lambda^{\phantom{\lambda}\nu}+R\,R_\mu^{\phantom{\mu}\nu} \nonumber\\
   & & -\frac{1}{4}\delta_\mu^\nu\left(R_{\lambda\delta}^{\phantom{\lambda\delta}\sigma\rho}R_{\sigma\rho}^{\phantom{\sigma\rho}\lambda\delta}-4R_\lambda^{\phantom{\lambda}\delta}R_\delta^{\phantom{\delta}\lambda}+ R^2\right)\label{gp5},
\end{eqnarray}
one can find the Gauss-Bonnet contribution to the Einstein tensor. Then it is convenient to represent the tensor components of $h_{\mu\nu}$ with respect to the transformation law under rotations on the $(D-2)$-sphere. The linear perturbations, then, can be classified into \emph{tensor}, \emph{vector}, and \emph{scalar} types, each of which can be treated independently from the others. After a lot of algebra and separation of variables \cite{Takahashi:2010ye}, the perturbation equations can be reduced to a number of second-order master differential equations with some effective potentials,
\begin{equation}\label{gp9}
\left(\frac{\partial^2}{\partial t^2}-\frac{\partial^2}{\partial r_*^2}+V_i(r_*)\right)\Psi(t,r_*)=0,
\end{equation}
where $r_*$ is the tortoise coordinate,
\begin{equation}
dr_*\equiv \frac{dr}{f(r)}=\frac{dr}{1-r^2\psi(r)},
\end{equation}
and $i$ stands for $t$ (tensor), $v$ (vector), and $s$ (scalar) perturbations.
The explicit forms of the effective potentials $V_s(r)$, $V_v(r)$, and $V_t(r)$ are given \cite{Takahashi:2010ye}
\begin{eqnarray}\label{potentials}
V_t(r)&=&\frac{\ell(\ell+n-1)f(r)T''(r)}{(n-2)rT'(r)}+\frac{1}{R(r)}\frac{d^2}{dr_*^2}\Biggr(R(r)\Biggr),\\\nonumber
V_v(r)&=&\frac{(\ell-1)(\ell+n)f(r)T'(r)}{(n-1)rT(r)}+R(r)\frac{d^2}{dr_*^2}\Biggr(\frac{1}{R(r)}\Biggr),\\\nonumber
V_s(r)&=&\frac{2\ell(\ell+n-1)}{nr^2B(r)}\frac{d}{dr_*}\Biggr(rB(r)\Biggr)+B(r)\frac{d^2}{dr_*^2}\left(\frac{1}{B(r)}\right),
\end{eqnarray}
where $\ell=2,3,4,\ldots$ is the multipole number and
\begin{eqnarray}
\nonumber
T(r)= r^{n-1}\frac{dW}{d\psi}&=&\frac{nr^{n-1}}{2}\Biggr(1+\alpha(n - 1) (n - 2)\psi(r)\Biggr),\\
\nonumber
R(r)=r\sqrt{T'(r)},&&\!\!\!\!\!\!B(r)=\frac{2(\ell-1)(\ell+n)-nr^3\psi'(r)}{r\sqrt{T'(r)}}T(r).
\end{eqnarray}

\section{Characteristic integration}\label{sec:timeint}

In our case the dynamical wave equation has a cumbersome form. Therefore, proving of (in)stability analytically is a difficult task, so that numerical analysis of the quasinormal spectrum must be undertaken instead. The most straightforward way to achieve this is to use the integration of the master equation in the time domain which takes into consideration contributions from all the modes.

We shall use here the discretization scheme proposed by Gundlach, Price, and Pullin \cite{Gundlach:1993tp}. This method was used for calculation of quasinormal modes in a great number of works  \cite{Konoplya:2011qq}. Comparisons of the time-domain numerical data with the accurate frequency-domain calculations show excellent agreement not only in cases when a black hole is stable, but also near the onset of instability (see for example \cite{Konoplya:2014lha}).  Rewriting (\ref{gp9}) in terms of the light-cone coordinates $du=dt-dr^*$ and $dv=dt+dr^*$, one finds
\begin{equation}\label{timedomain}
4\frac{\partial^2\Psi}{\partial u\partial v}=-V_i(u-v)\Psi.
\end{equation}

The discretization scheme has the following form
\begin{eqnarray}
\Psi(N)&=&\Psi(W)+\Psi(E)-\Psi(S)\label{ci1}\\
&&-\frac{\Delta^2}{8}V_i(S)\left[\Psi(W)+\Psi(E)\right]+\mathcal{O}(\Delta^4),\nonumber
\end{eqnarray}
where $N$, $M$, $E$, and $S$ are the points of a square in a grid with step $\Delta$ in the discretized $u$-$v$ plane: $S=(u,v)$, $W=(u+\Delta,v)$, $E=(u,v+\Delta)$, and $N=(u+\Delta,v+\Delta)$. With the initial data specified on two null surfaces $u = u_0$ and $v = v_0$, we are able to find values of the function $\Psi$ at each of the points of the grid. Since quasinormal modes and the asymptotical behavior of perturbations do not depend on initial conditions (as confirmed by several numerical simulations), we shall consider the Gaussian wave initial data on the $v$-axes (see \cite{Konoplya:2011qq} for more details).

This discretization scheme requires a number of operations which is proportional to $\Delta^{-2}$, what implies that the corresponding accumulated error is $\mathcal{O}(\Delta^2)$. By decreasing the step $\Delta$ for the same initial data, we check the convergence of the integration scheme. Here we present the resulting profiles for the sufficiently small step $\Delta$, such that its further decreasing does not change the time-domain picture. Another source of error comes from numerical truncations in computations. In order to check the stability of the algorithm, we compare the profiles found with the floating-point arithmetics of different precision. In particular, we observe that the single-precision (32-bits) arithmetics is sufficient for our computations. We have also compared the obtained time-domain profiles for $\Lambda=0$ with those obtained with the high-precision code in \cite{Konoplya:2008ix} and found that the difference is smaller than the discretization-scheme error order.

\begin{figure}
\resizebox{\linewidth}{!}{\includegraphics*{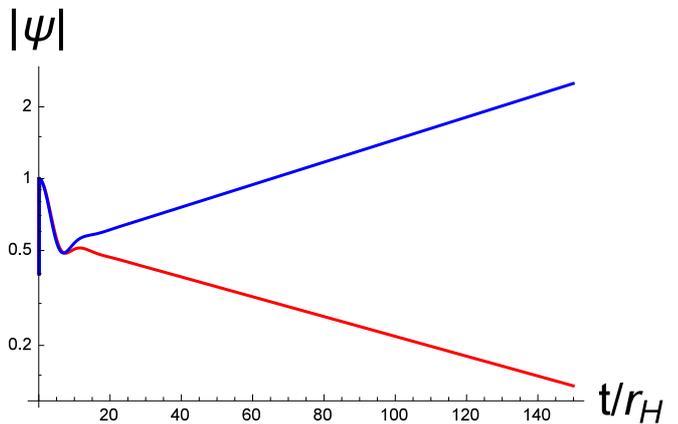}}
\caption{Nonoscillatory temporal profiles for scalar perturbations for $\alpha=0.5r_H^2$. The stable (red) and unstable (blue) profiles correspond to $\Lambda=0.8/r_H^2$ and $\Lambda=0.9/r_H^2$.}\label{profile-1}
\end{figure}

In order to catch the threshold of instability, we had to obtain a great number of time-domain profiles for various values of parameters. A typical damped (stable) and growing (unstable) time-domain profiles near the threshold of instability are shown on Fig.~\ref{profile-1}.

\section{(In)stability of Gauss-Bonnet-de Sitter black holes}\label{sec:res}

Here we shall consider the results of our numerical time-domain integration in terms of stability or instability of Gauss-Bonnet-de Sitter black holes at various values of parameters. We observed instability in scalar and tensor channels of perturbations, while the vector sector showed no growing time-domain profiles.

\subsection{Instability of tensor-type perturbations}

The effective potential for the tensor-type perturbations has a negative gap outside the black hole, near its event horizon. Although, intuitively, increasing of the multipole number $\ell$ should lead to a higher barrier of the effective potential, this is not the only effect that comes from increasing of $\ell$. The higher $\ell$ is, the deeper the negative gap, so that, quite conterintuitively, the higher $\ell$ are more unstable.
Therefore, in order to determine the instability region in the tensor channel, we have to consider the limit $\ell=\infty$, corresponding to the most unstable solution. In order to distinguish this instability, which develops at higher $\ell$, we shall call it the \emph{eikonal instability}, emphasizing the fact that the regime of geometrical optics $\ell =\infty$ corresponds to the most unstable solution.

In \cite{Reall:2014pwa}, it was shown that for Lovelock theories the negative-energy bound state in the $\ell=\infty$ limit exists in the region (outside the black-hole horizon) for which the initial conditions \emph{on spacelike surfaces} cannot be consistently imposed. Although, for the initial conditions on the \emph{null surfaces} considered in the present paper, the solutions of the equation (\ref{timedomain}) are well-defined (yet summing over $\ell$ is divergent) and correspond to unstable time-domain profiles. Thus, the parametric region of what we call ``the eikonal instability'' coincides exactly with the region where the perturbation equations become nonhyperbolic. However, for practical purposes, within a linear approximation it does not matter whether the configuration is unstable or has the region where the initial conditions are ill-defined. In the regime of small perturbations, we are unable to tell what occurs once we observe a profile infinitely growing in time. However, we are able to demonstrate the parametric region in which the black holes are stable and find the onset of instability (at which he damping rate approaches zero). As the problem of non-hyperbolicity of the perturbation equations occurs only when the ``eikonal instabilty'' takes place, this gives us an additional reason to distinguish the ``eikonal instability'' from the other type of instability which is not related to this problem.

\begin{figure}
\resizebox{\linewidth}{!}{\includegraphics*{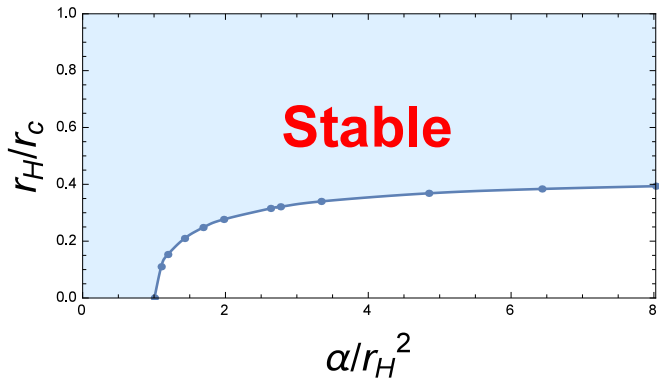}}
\caption{Stability and instability regions for tensor-type gravitational perturbations in $(D=6)$-dimensional Gauss-Bonnet-de Sitter spacetimes. This is the eikonal instability with the $\ell=\infty$ most unstable mode.}
\label{Tensor-6}
\end{figure}

From Fig.~\ref{Tensor-6}, we can see that small black holes are always unstable against tensor-type perturbations in $D=6$ spacetimes.
For higher $D$, as well as for $D=5$, tensor-type perturbations show no instability.
Apparently, the critical value of the fraction $r_H/r_C$, corresponding to the threshold of instability at a given $\alpha$, approaches to some constant for large $\alpha$. In this sense, there should exist a minimum value of $r_H/r_C$, designated as $\mu_c$, such that black holes with $r_H/r_C>\mu_c$ are stable against tensor perturbations in Gauss-Bonnet gravity for any values of the coupling constant $\alpha$.

\subsection{Instabilities of scalar-type perturbations}

\begin{figure*}
\resizebox{\linewidth}{!}{\includegraphics*{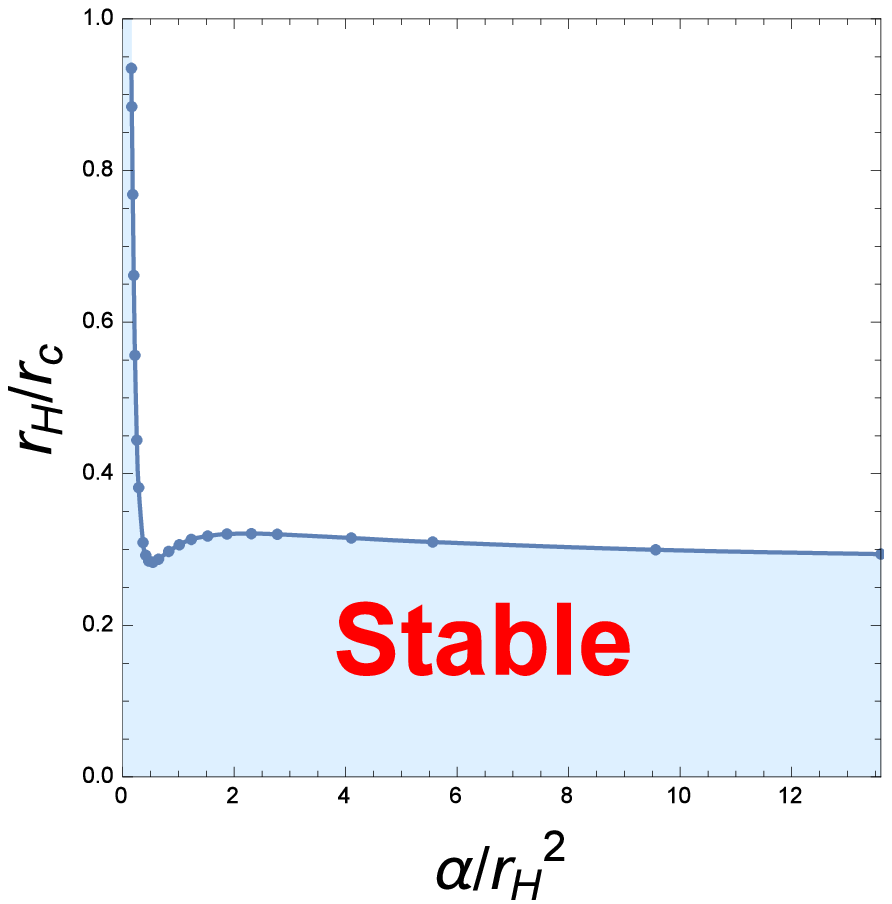}\includegraphics*{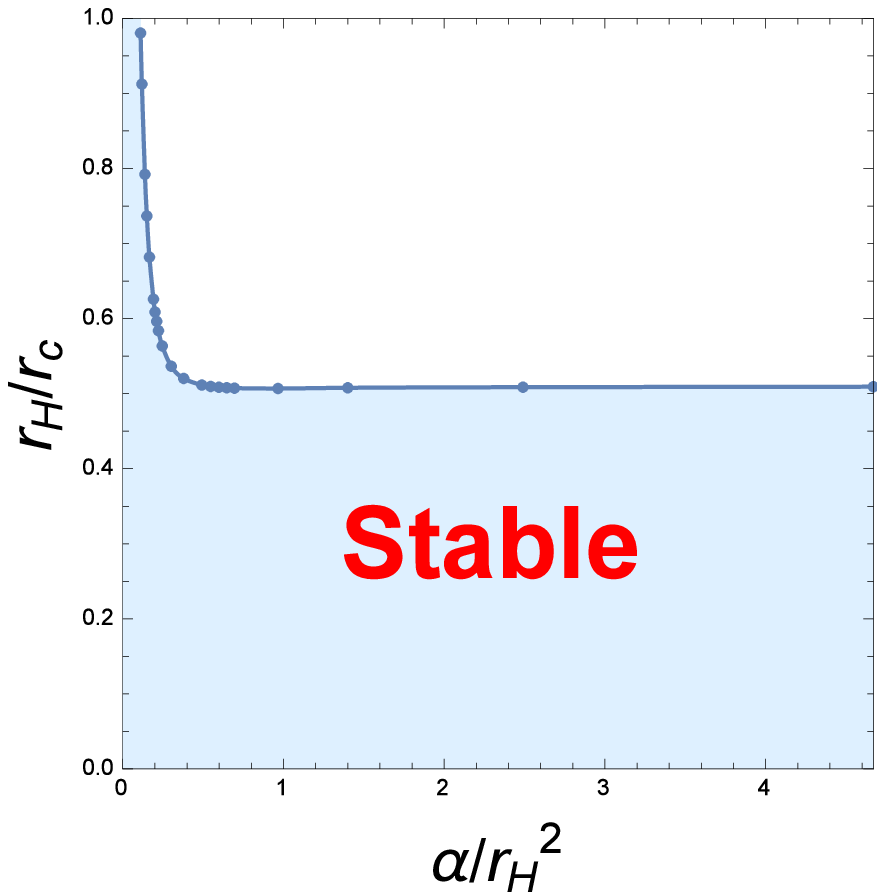}\includegraphics*{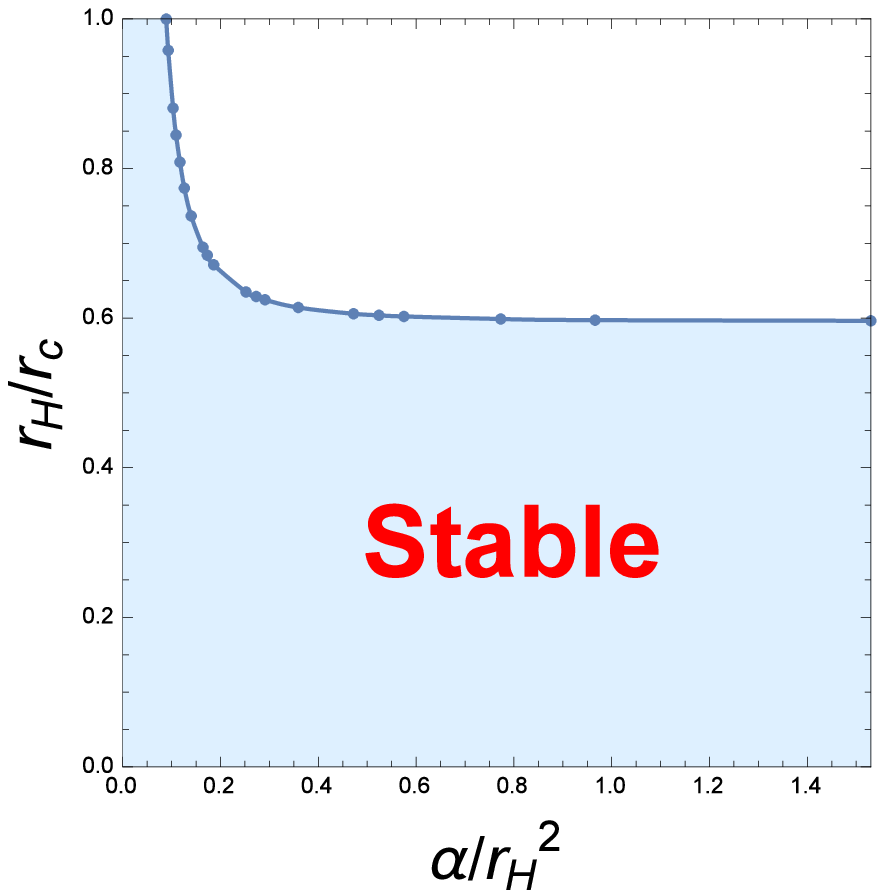}}
\caption{Stability and instability regions for scalar-type gravitational perturbations in $D=6$, $D=7$, and $D=8$ (from left to right).}\label{InstabilityS}
\end{figure*}

\begin{figure}
\resizebox{\linewidth}{!}{\includegraphics*{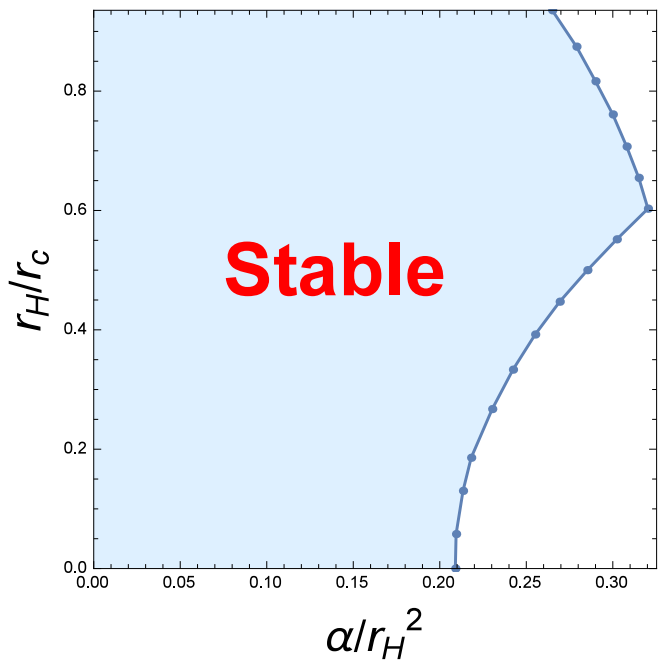}}
\caption{Stability and instability regions for scalar-type gravitational perturbations in 5 dimensions. Upper right corner corresponds to the $\Lambda$-instability, while the lower right corner corresponds to the eikonal instability. The overlap of regions of both types of instability produce the instability region for $D=5$ case.}\label{InstabilityS5}
\end{figure}

Analyzing the time-domain profiles, we see that unboundedly growing perturbations, signifying the instability, correspond to the nonoscillatory regime near the threshold of instability. This observation is in agreement with the statement, proved in \cite{Konoplya:2008yy}, that \emph{unstable modes cannot be oscillatory} when perturbing spherically symmetric static black holes.

In the gravitational perturbations of the scalar type we observe the new instability for all $D \geq 5$. This instability occurs for sufficiently large value of the GB-coupling and cosmological constant (Fig.~\ref{InstabilityS}). Unlike the eikonal instability it occurs for the lowest multipole number $\ell=2$ and, therefore, is not related to the above mentioned nonhyperbolicity problem of perturbation equations. Since this instability does not take place for asymptotically flat Gauss-Bonnet black holes we shall call it, for briefness, \emph{$\Lambda$-instability}.

Scalar-type gravitational perturbations of the five-dimensional black holes are different from higher-dimensional cases, because only in $D=5$ spacetimes both types of instability, the eikonal one and the $\Lambda$-instability, take place. From Fig.~\ref{InstabilityS5}, we can see that there are two regimes of instability:
\begin{itemize}
\item \emph{Small} $D=5$ black holes are unstable once the GB-coupling is larger than some critical value (for a given $r_H/r_C$) $\alpha=\alpha_{crit}$ The instability region is dominated by the eikonal $\ell=\infty$ regime.
\item \emph{Large} $D=5$ black holes are unstable for values of the coupling $\alpha$ above some critical (at a given $r_H/r_C$). This instability happens only $\ell=2$ modes.
\end{itemize}

Close to $\Lambda=1.6r_H^{-2}$ $(r_H=0.6r_C)$ both types of instability ``merge'' in such a way, that there seems to be no values of $\mu_c=r_H/r_C$ for which the black hole is stable for \emph{arbitrary GB-coupling}, like it happens in $D=6$ tensor-type mode.

\begin{figure}
\resizebox{\linewidth}{!}{\includegraphics*{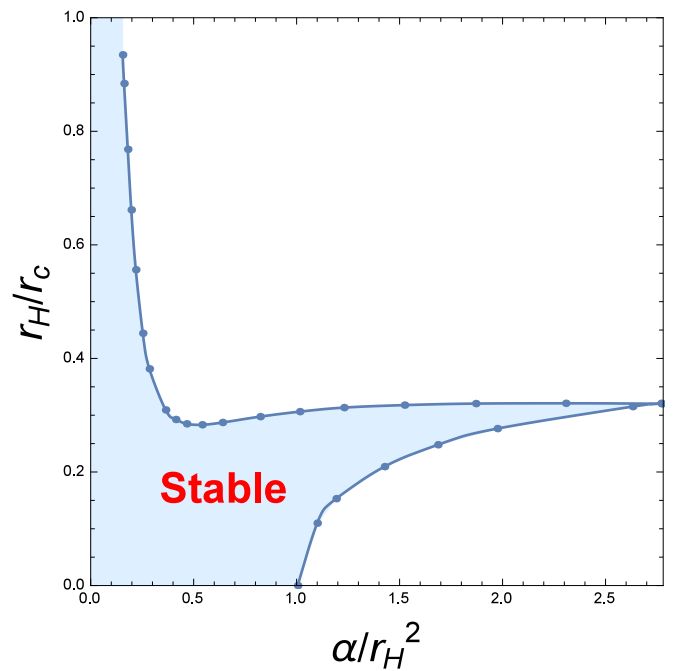}}
\caption{Stability and instability regions for gravitational perturbations in $D=6$ as overlap of the $\Lambda$-instability in the scalar channel and eikonal instability in the tensor channel.} \label{InstabilityS6}
\end{figure}

\begin{figure}
\resizebox{\linewidth}{!}{\includegraphics*{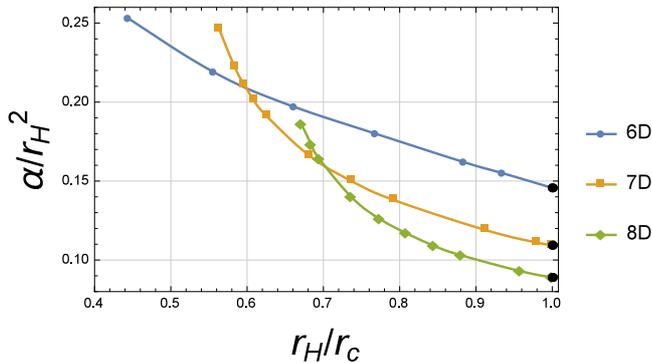}}
\caption{Scalar type of gravitational perturbations: Critical value of $\alpha$, corresponding to the threshold of instability as function of $r_H/r_C$ in dimensions $6$, $7$, and $8$.}\label{alpha-min}
\end{figure}

\begin{table}[ht!]
  \centering
  \begin{tabular}{|c|c|c|}
    \hline
    D=6 & D=7 & D=8\\
    \hline
    \begin{tabular}{c|c}
     $r_H/r_c(\Lambda\,r_H^2)$ & $\alpha/r_H^2$\\
    \hline
    0.662(4) & 0.197 \\
    0.768(5) & 0.180 \\
    0.884(6) & 0.162 \\
    0.935(6.4) & 0.155\\
    1 (6.876) & 0.146
    \end{tabular}
     &
     \begin{tabular}{c|c}
     $r_H/r_c(\Lambda\,r_H^2)$ & $\alpha/r_H^2$\\
     \hline
     0.737(9) & 0.150\\
    0.792(10) & 0.138\\
    0.912(12) & 0.119\\
     0.980(13) & 0.111\\
     1 (13.27) & 0.109
    \end{tabular}
     &
    \begin{tabular}{c|c}
     $r_H/r_c(\Lambda\,r_H^2)$ & $\alpha/r_H^2$\\
    \hline
    0.845(19) & 0.109\\
    0.881(20) & 0.103\\
    0.957(22) & 0.093\\
    0.999(23) & 0.089\\
    1 (23.01) & 0.089
    \end{tabular}\\
    \hline
  \end{tabular}
  \caption{Scalar type of gravitational perturbations: Critical values of $\alpha$ corresponding to the threshold of instability.}
  \label{tab:interpolateval}
\end{table}

From Fig.~\ref{InstabilityS} we can see that sufficiently small black holes are always stable against scalar perturbations, because there exists a $\mu_c$ such that for $r_H<\mu_c\,r_C$, Gauss-Bonnet black holes are stable. The $D=6$ case is slightly different from $D=7, 8,\ldots$ because the instability region is the overlap of $\Lambda$-instability in the scalar channel and eikonal instability in the tensor channel (see Fig.~\ref{InstabilityS6}). $D\geq5$ black holes are apparently stable for all $r_H/r_C$, including the extremal limit $r_H=r_C$ once $\alpha$ is less than some minimum value, which depends on the number of spacetime dimensions $D$. This minimal value decreases as $D$ grows, what can be seen in Fig.~\ref{alpha-min} and in the Table~\ref{tab:interpolateval}.

\subsection{Overlap of the instabilities in tensor- and scalar-type perturbations}

Here we shall review the results obtained for scalar, vector, and tensor types of gravitational perturbations. First of all, we would like to remind that the vector type of perturbations does not show any instability for all $D$. Then, $D =7, 8, \dots$ black holes have only the $\Lambda$-instability in the scalar type of gravitational perturbation. The regions of this instability are shown on Figs.~\ref{InstabilityS}~and~\ref{alpha-min}. $D=5$ and $D=6$ spacetimes have more complicated regions of instability determined by \emph{the combination of the $\Lambda$-instability and eikonal instability}: in the case of $D=5$ the $\Lambda$-instability in the scalar channel combines with the eikonal instability in the same channel (Fig.~\ref{InstabilityS5}), while for $D=6$ the $\Lambda$-instability in the scalar channel combines with the eikonal instability in the tensor channel (Fig.~\ref{InstabilityS6}). The black hole instability region is a combination of the instability regions of all types of gravitational perturbations. Accordingly, the black-hole stability region is the overlap of the corresponding stability regions. The summary of the (in)stabilities are represented also in the Table~\ref{tab:inst}.

\begin{table}
\begin{tabular}{|c|c|c|}
  \hline
  $D$ & $\Lambda$-instability & eikonal instability \\
  \hline
  5 & scalar-type ($\ell =2$) & scalar-type\\
  \hline
  6 & scalar-type ($\ell =2$) & tensor-type\\
  \hline
  7 & scalar-type ($\ell =2$) & \\
  \hline
  8 & scalar-type ($\ell =2$) & \\
  \hline
\end{tabular}

\caption{Summary of instabilities of Einstein-Gauss-Bonnet-de Sitter black holes: each type of instability implies its parametric region. For the eikonal instability this region expands as $\ell$ increases, so that the instability region in this case corresponds to the limit $\ell\rightarrow\infty$.}\label{tab:inst}
\end{table}

\section{Quasinormal modes of Gauss-Bonnet-de Sitter black holes and Hod's conjecture}\label{sec:Hod}

Several years ago Shahar Hod formulated an interesting proposal \cite{Hod:2006jw} stating that the \emph{damping rate of the fundamental quasinormal frequency} of any black hole in nature is constrained by the value of its Hawking temperature in a specific way. Namely, he argued that
\begin{equation}
|\im{\omega}| \leq \pi T_{H},
\end{equation}
where $T_H$ is the Hawking temperature. For static spherically symmetric black holes
\begin{equation}
T_{H} = \frac{\kappa_H}{2 \pi} = \frac{f'(r_H)}{4 \pi}\,.
\end{equation}
Using numerical and analytical results for quasinormal modes of four- and higher-dimensional Schwarzschild \cite{Konoplya:2003ii}, Schwazrschild-de Sitter \cite{Zhidenko:2003wq} and Schwarzschild-anti-de Sitter black holes \cite{Horowitz:1999jd}, Hod illustrated that his inequality is fulfilled for asymptotically flat black holes as well as  for nonasymptotically flat ones. The arguments were based on semiclassical consideration and thermodynamic ideas.

\begin{figure*}
\resizebox{\linewidth}{!}{\includegraphics*{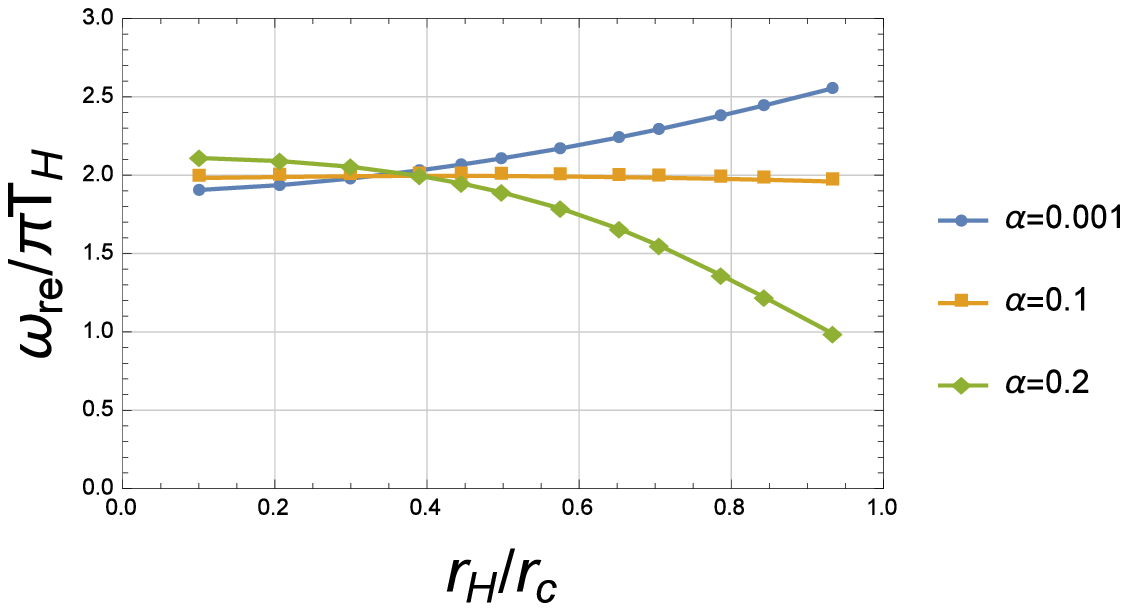}\includegraphics*{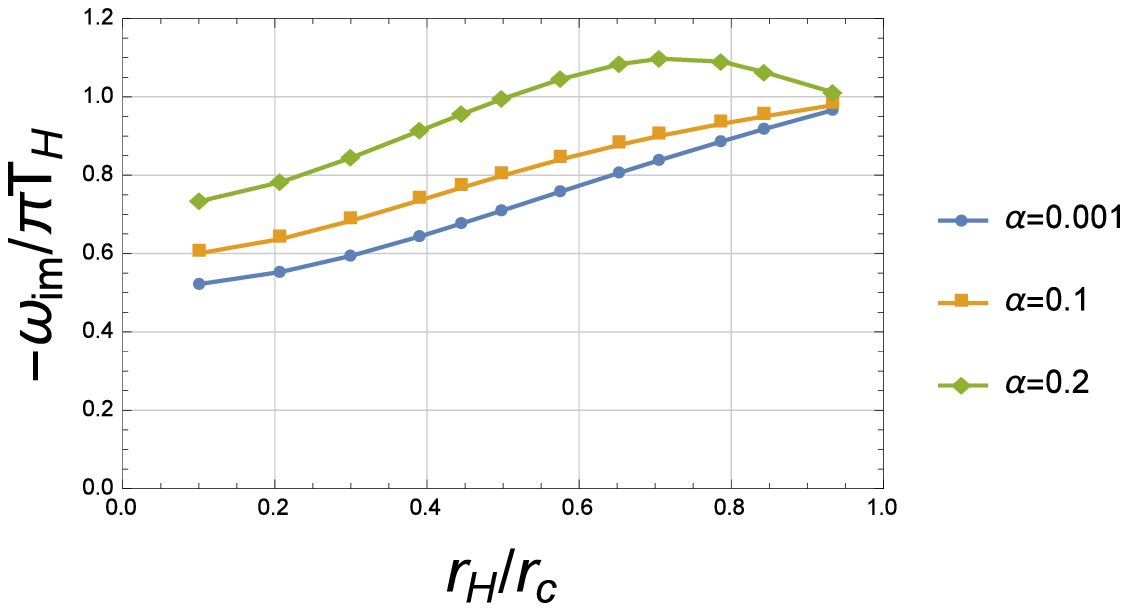}}
\caption{Variation of the real (left panel) and imaginary (right panel) parts of the dominant quasinormal mode as functions of $r_H/r_C$ for scalar-type perturbations in 5 dimensions.}\label{QNMT}
\end{figure*}

It is tempting to understand whether Hod's proposal is valid for more general black hole solutions. In our case, we choose the $D=5$ Gauss-Bonnet-de Sitter black hole in the range of parameters which corresponds to a gravitationally stable configuration. The breakdown of the inequality at the onset of instability would be a perfect proof of the proposal. However, from Fig.~\ref{QNMT} we noticed that the lowest mode of the scalar type of gravitational perturbations has the imaginary part $\im{\omega}$ for which

\begin{figure}
\resizebox{\linewidth}{!}{\includegraphics*{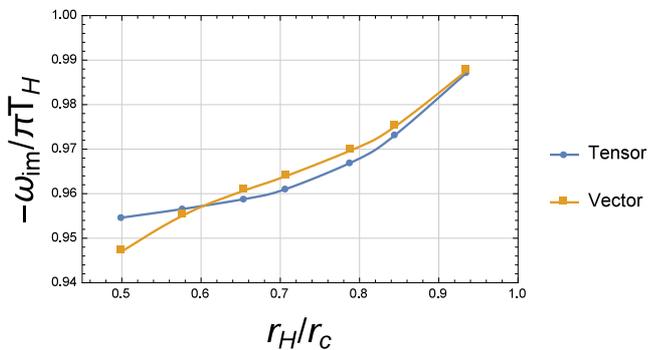}}
\caption{Variation of the imaginary part of the dominant quasinormal mode with respect to $r_H/r_C$ for vector- and tensor-type perturbations in 5 dimensions for $\alpha=0.2r_H^2$.}\label{QNMTvt}
\end{figure}

\begin{equation}
|\im{\omega}| > \pi T_{H},
\end{equation}
for $\alpha =0.2r_H^2$ in the range $0.5\lesssim r_H/r_C \lesssim 0.94$, which is inside the stability region according to Fig.~\ref{InstabilityS5}.
Nevertheless, vector and tensor perturbations in the stable sector (for $\alpha=0.2r_H^2$), unlike scalar perturbations, do not break down the conjecture, as we can see in Fig.~\ref{QNMTvt}.

Thus, if we admit that it is possible to ``create'' perturbation in such a way that only the scalar type of gravitational perturbation would be excited, while vector- and tensor-type gravitational modes would have zero excitation factors, the counterexample for this proposal would be provided. Even though all three types of perturbations can be treated completely separately within the linear approximation, we do not believe that it would be possible to prepare such process of ``purely scalar-type gravitational perturbation'' in a real process. The main reason for this is that such an ideal separation of gravitational perturbations into three independent channels (scalar, vector, and tensor) is impossible taking into account the essentially nonlinear nature of the real perturbation process. Such a separation into three channels would also be impossible even in the linear approximation: Although a classical black hole preserves all the geometrical symmetries given by a set of Killing vectors, tiny quantum fluctuations definitely break down the exact symmetries. In other words, if due to specially prepared initial perturbations only one of the channels (scalar) is excited, then the other two channels (vector and tensor) inevitably acquire some, whatever small, but nonzero, excitations. Therefore, by showing that at least some of the types of gravitational perturbations obey Hod's proposal, we rather confirm the proposal than disprove it. This certainly does not mean that counter-examples cannot be found in the future.

We can also see that the higher the value of the $\alpha$-coupling, the quicker the Hawking temperature grows with respect to the oscillation frequency. The decay rate of oscillation, $\im{\omega}/\pi\,T_H$, reaches a constant value when approaching the extremal state. Indeed, the imaginary part of the QNM approaches zero for extremal black holes, so that $\im{\omega}\propto T_H\approx0$ for $r_H\approx r_C$.

\section{Conclusion}

Here we have performed a thorough analysis of gravitational quasinormal spectrum of asymptotically de Sitter black holes in Einstein-Gauss-Bonnet theory. Usage of the time-domain integration allowed us to take into consideration contributions of all the modes in the signal and, thereby, to judge about the (in)stability of the black hole (what would be much more difficult to do by working in the frequency domain). Gravitational perturbations are known to be reduced to the independent master equations for scalar, vector, and tensor types relatively the rotation group  on $(D-2)$-sphere. It has been shown that the scalar channel of the gravitational perturbation has a new kind of instability at sufficiently large values of the cosmological constant $\Lambda$, which we called ``the $\Lambda$-instability'', because it does not take place for asymptotically flat spacetimes. It is possible that this instability has a similar origin as the instability of the higher-dimensional Reissner-Nordstr\"om-de Sitter black holes \cite{Konoplya:2008au}, though there are apparent distinctions between these two instabilities: Reissner-Nordstr\"om-de Sitter black holes are unstable in $(D>6)$-dimensional spacetimes only at relatively large values of the electric charge, while black holes in the Gauss-Bonnet theory are unstable even being neutral and in $5$ and $6$ dimensions as well. In addition, we have found that scalar and tensor channels also have instabilities owing to the nonzero Gauss-Bonnet coupling. This instability occurs at high multipole numbers $\ell$ and, therefore, was called ``the eikonal instability''.

It was demonstrated that the quasinormal frequencies of the scalar type of gravitational perturbations do not obey Hod's inequality and the lowest mode in this channel of perturbation has higher damping rate than the one prescribed by the proposal. However, the other two channels, vector and tensor, have lower-lying modes what, thereby, confirms Hod's proposal. Apparently, it would be impossible to create the process of perturbation which would excite only the scalar channel and leave unperturbed the other two. If such a ``selective'' perturbation could be prepared or even theoretically modeled, Hod's proposal would be violated.

In the forthcoming papers \cite{USprogress}, we shall complete the investigation of the (in)stabilities of black holes in Gauss-Bonnet gravity and consider perturbations of the electrically charged asymptotically flat, de Sitter, and anti-de Sitter black holes in Gauss-Bonnet gravity. Some of these cases require application of different numerical approaches and deserve separate consideration.

\section*{Acknowledgments}
M.~A.~C.~was supported by Coordena\c{c}\~ao de Aperfei\c{c}oamento de Pessoal de N\'ivel Superior (CAPES).
R.~A.~K.~would like thank Andrei Starinets for useful discussions.
A.~Z.~was supported by Conselho Nacional de Desenvolvimento Cient\'ifico e Tecnol\'ogico (CNPq).

\end{document}